\documentclass[10pt,conference]{IEEEtran}
\IEEEoverridecommandlockouts
\usepackage{cite}
\usepackage{amsmath,amssymb,amsfonts}
\usepackage{algorithmic}
\usepackage{graphicx}
\usepackage{textcomp}
\usepackage{xcolor}
\usepackage{graphicx}
\usepackage{hyperref}       
\usepackage{url}            
\usepackage{booktabs}       
\usepackage{amsfonts}   
\usepackage{amsmath}
\usepackage{nicefrac}       
\usepackage{microtype}      
\usepackage{lipsum}
\usepackage{multirow}
\usepackage{subcaption}
\usepackage{array}
\usepackage{enumitem}
\usepackage[ruled, vlined, linesnumbered]{algorithm2e}
\usepackage[english]{babel}
\usepackage{multirow}
\usepackage{color}
\usepackage{caption}

\usepackage[font=small,skip=0pt]{caption}

\usepackage[belowskip=-15pt,aboveskip=0pt]{caption}

\usepackage{titlesec}
\titlespacing*{\section}{-4pt}{\dimexpr\baselineskip-4pt}{\baselineskip}
\def\BibTeX{{\rm B\kern-.05em{\sc i\kern-.025em b}\kern-.08em
    T\kern-.1667em\lower.7ex\hbox{E}\kern-.125emX}}
\makeatletter

\begin{document}
\title{Enhanced Graph Neural Networks with Ego-Centric Spectral Subgraph Embeddings Augmentation}

\author{\IEEEauthorblockN{Anwar Said\textsuperscript{\dag}, Mudassir Shabbir\textsuperscript{\dag}\textsuperscript{\S}, Tyler Derr\textsuperscript{\dag}, Waseem Abbas\textsuperscript{\ddag}, Xenofon Koutsoukos\textsuperscript{\dag}
\thanks{Anwar~Said, Tyler~Derr and Xenofon~Koutsoukos are with the Computer Science Department at the Vanderbilt University, Nashville, TN. Emails: \{anwar.said,tyler.derr,xenofon.koutsoukos\}@vanderbilt.edu.} 
\thanks{Waseem~Abbas is with the Systems Engineering Department at the University of Texas at Dallas, Richardson, TX. Email: waseem.abbas@utdallas.edu}
\thanks{Mudassir~Shabbir is with the Computer Science Department at Information Technology University, Lahore, Pakistan and with Vanderbilt University, Nashville, TN, USA. Email: mudassir.shabbir@itu.edu.pk}
\thanks{22nd IEEE International Conference on Machine Learning and Applications 2023}
}

\IEEEauthorblockA{
\textsuperscript{\dag}Vanderbilt University, Nashville, TN, USA\\
\textsuperscript{\ddag}University of Texas at Dallas, Richardson, TX, USA\\
\textsuperscript{\S}Information Technology University, Lahore, Pakistan\\
}
}

\maketitle

\begin{abstract}
Graph Neural Networks (GNNs) have shown remarkable merit in performing various learning-based tasks in complex networks. The superior performance of GNNs often correlates with the availability and quality of node-level features in the input networks. However, for many network applications, such node-level information may be missing or unreliable, thereby limiting the applicability and efficacy of GNNs. To address this limitation, we present a novel approach denoted as Ego-centric Spectral subGraph Embedding Augmentation (ESGEA), which aims to enhance and design node features, particularly in scenarios where information is lacking. Our method leverages the topological structure of the local subgraph to create topology-aware node features.  
The subgraph features are generated using an efficient spectral graph embedding technique, and they serve as node features that capture the local topological organization of the network. The explicit node features, if present, are then enhanced with the subgraph embeddings in order to improve the overall performance. ESGEA is compatible with any GNN-based architecture and is effective even in the absence of node features. We evaluate the proposed method in a social network graph classification task where node attributes are unavailable, as well as in a node classification task where node features are corrupted or even absent. The evaluation results on seven datasets and eight baseline models indicate up to a 10\% improvement in AUC and a 7\% improvement in accuracy for graph and node classification tasks, respectively.
\end{abstract}

\begin{IEEEkeywords}
Graph Neural Networks, Subgraph Spectral Embeddings, Graph Descriptors, Abnormal Features 
\end{IEEEkeywords}

\vspace{-1ex}
\section{Introduction}\vspace{-0.15in}
\label{introduction}

Graph representation learning has proved crucial to several real-world applications, including drug discovery \& development \cite{stokes2020deep}, weather and traffic forecasting \cite{derrow2021eta}, recommendation in e-commerce \cite{wu2020graph},  combinatorial optimization \cite{cappart2021combinatorial}, etc. In the past few years, there has been a surge of interest in designing graph neural networks (GNNs), which are powerful tools for learning from graph-structured data \cite{kipf2016semi,hamilton2017inductive,choma2018graph}. GNNs have attained state-of-the-art performance on a variety of downstream Machine Learning (ML) tasks, such as node classification, graph classification, graph regression, and link prediction \cite{ying2021transformers,said2021netki}. For example, predicting the toxicity or property of molecules, item recommendations in an e-commerce website, and identifying users' communities in social networks \cite{wu2020comprehensive,derr2020epidemic}.  

\begin{figure}[!t]
\centering
\begin{minipage}{.255\textwidth}
  \hspace*{-1ex}\includegraphics[width=1.0\textwidth]{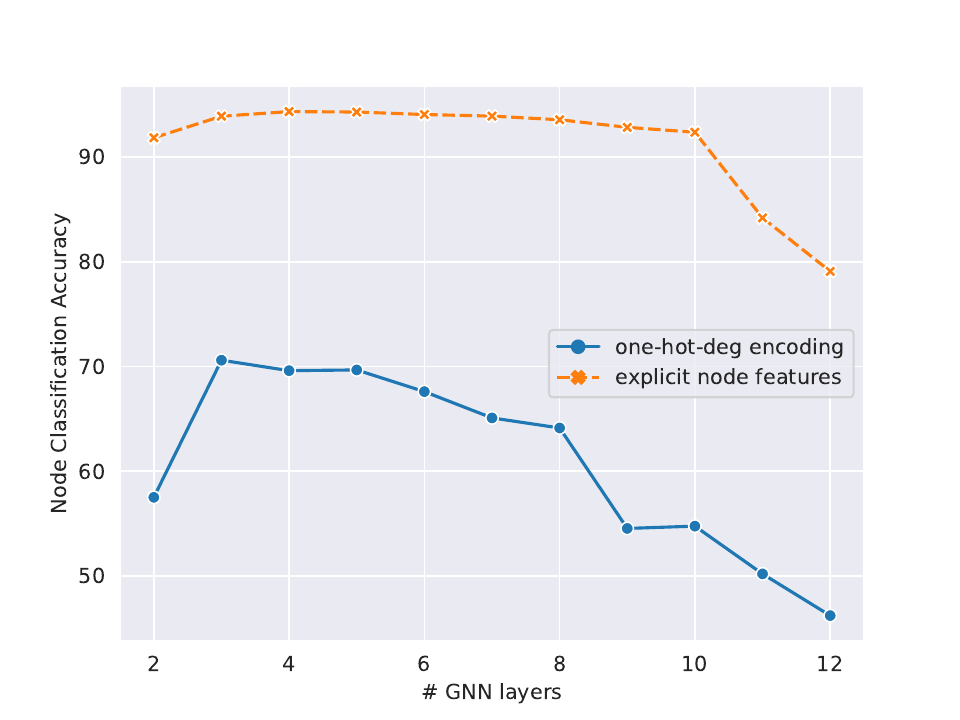}
\end{minipage}%
\begin{minipage}[width=0.255\textwidth]{.245\textwidth}
  \hspace*{-1ex}\includegraphics[width=1.0\textwidth]{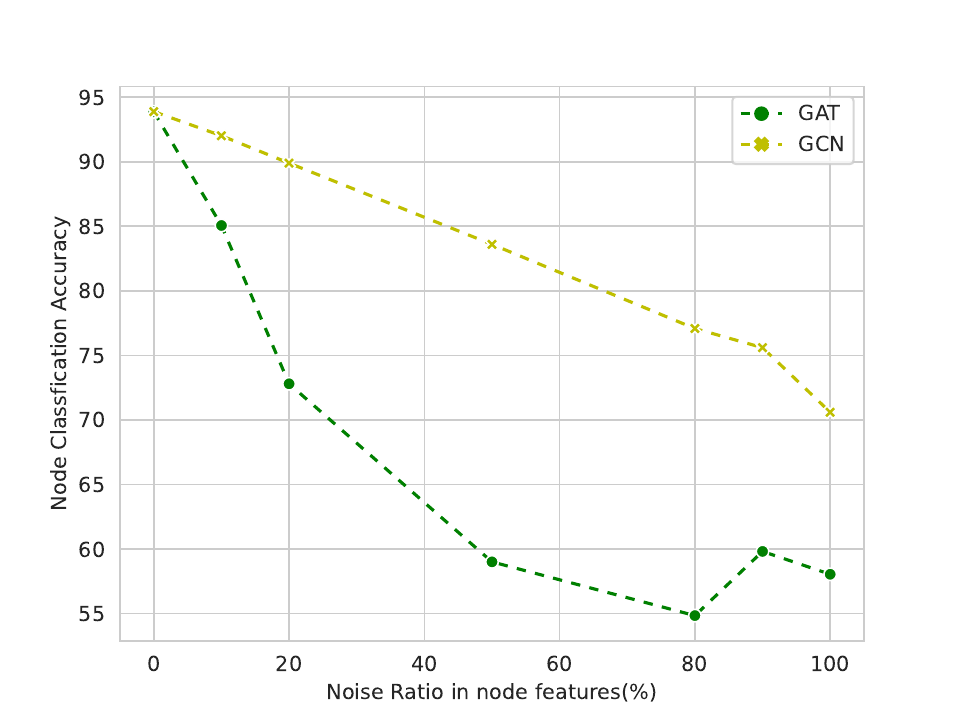}
\end{minipage}
\vspace{0.05in}
\caption{(a) GCN's performance on the Facebook dataset \cite{rozemberczki2021multi} using original features and one-hot-degree encoding. (b) performance with varying abnormality/noise ratios in the node features. }
\label{fig:examples}
\vspace{0.23in}
\end{figure}

One key advantage of GNNs over manually engineered embeddings is the ability to learn a correlation between information specific to a node and its \textit{global} position in the network. Explicit node features often play an integral role in the performance of GNNs since they encode valuable distinguishing criteria about the entities. For instance, in molecular network data, node features provide crucial information about the chemical nature of the element. Similarly, in citation networks, node features provide textual information on represented publications and contribute significantly to the model's overall performance. When this node-level information is unavailable, current methods use ad hoc techniques such as random vectors or vectors of ones. Numerous recent approaches also use one-hot degree encoding \cite{xu2018powerful}. However, this is rarely a suitable substitute for the explicit node's features. In Figure \ref{fig:examples} (a), we illustrate this by comparing the performance of explicit node features and one-hot-degree encoding on Graph Convolutional Networks (GCN) \cite{kipf2016semi} with varying numbers of layers. We observe up to 25\% improvement in the results for the explicit features compared to the degree encoding. These results evince that one-hot degree encoding significantly degrades the model's performance, necessitating the use of robust techniques to generate expressive node embeddings in graph lacking node features. Specifically, the design of a framework that produces topology-aware node features in situations when node features are unavailable is one goal of this study.

The node features could be missing or abnormal for a number of reasons, such as privacy concerns, human or machine error, adversarial error, and incomplete data entry \cite{rossi2021unreasonable}. For instance, in a social network, users may not have completed their profile information, resulting in missing user features. Similarly, not all items in a co-purchase network may have a complete description associated with them. In a transportation network, traffic information coming from sensors may be noisy due to complex dynamics, leading to abnormal node features. 
We illustrated the effect, in Figure~\ref{fig:examples}~(b) by running Graph Attention Network (GAT) \cite{velivckovic2017graph}, and GCN on varying ratios of random Gaussian noise in the node features on Facebook dataset \cite{rozemberczki2021multi}. We 
observe a significant decline in performance when noise is injected. 
Thus, the second objective of this study is to design a mechanism 
resilient to abnormal node features.

We observe that node features in many social networks originate the local subgraph topology. For example, keeping the number of followers as a node feature in a Facebook-page network reflects the in-degree of the node. At the same time, we can gather sufficient information of 
the graph's structure from its subgraphs. The 
reconstructibility conjecture, which holds true for many graph families, asserts that a graph can be exactly constructed from a collection of its (single) vertex-deleted subgraphs~\cite{bilinski2007reconstruction,bondy1977graph}. A vertex-deleted subgraph for a vertex $v$ is obtained by deleting $v$ from the graph. Even more, if $B_k(v)$ is the set of vertices at most $k$-hop from $v$, and $G_k(v)$ is the subgraph of $G$ obtained by deleting vertices in $B_k(v)$, then many graph properties can be inferred or a graph can be constructed exactly from a collection of such $G_k(v)$~\cite{levenshtein2008conjecture,kostochka2020reconstruction}. 

Therefore, we propose to use topology-aware subgraph embeddings that are based on the local topology of a network as features of a node. The proposed framework, denoted as Ego-centric Spectral subGraph Embeddings Augmentation (ESGEA) allows to design topology-aware node features using expressive graph embedding methods to enhance the existing node features. In applications where node features are unavailable, ESGEA provides a flexible way to produce node features that are expressive enough to obtain quality results. ESGEA consists of four modules: (a) ego-centric subgraph extraction, where $k-$hops local subgraphs are extracted for each node, (b), a spectral graph-embedding method is deployed to extract expressive graph representations on each subgraph, (c) feature augmentation module is proposed to augment features, and, (d) GNN learning module is provided to learn nodes/graph representations. The proposed approach is flexible to use any off-the-shelf graph-based embedding with any GNN-based architecture to design models for different applications. 
Unlike existing subgraph methods that learn node representations with nested subgraphs message passing, the proposed approach is novel in terms of bridging the gap with topology-aware graph descriptors to use with GNNs.
Moreover, ESGEA provides a flexible learning framework that can be customized to meet the desired goal. We offer the following contributions:
\vspace{-0.5ex}
\begin{itemize}[leftmargin=*]
    \item We introduce a topological feature augmentation method, Ego-centric Spectral subGraph Embedding Augmentation (ESGEA), that designs new or enhances corrupted/missing node features. \\\vspace{-2.25ex}
    \item We introduce a novel framework that offers 
    flexibility for
    graph representation pipelines by combining spectral graph-based embeddings with GNNs based on ESGEA.\\\vspace{-2.25ex}
    \item We evaluate the proposed framework in graph and node classification settings where node features are unavailable or corrupted (e.g., in the presence of varying amounts of noise) and show its effectiveness through extensive experiments.
\end{itemize}


The structure of this paper is as follows. Section 2 presents an overview of the related work. Section 3 provides the preliminaries and an introduction to a few definitions, while Section 4 details the methodology of the proposed approach. In Section 5, an evaluation of the proposed approach in both graph and node classification setting is provided. Finally, Section 6 concludes the paper and outlines potential future directions.   

\vspace{-0.5ex}
\section{Related Work}
\vspace{-0.95ex}
\label{sec:related-work}
Graph Neural Networks (GNNs) have made substantial advancements in learning representations of graph-structured data in recent years \cite{bronstein2017geometric}. GNNs essentially generalize end-to-end learning from regular grid data such as image, video, and text, to graph-structured data \cite{wu2020comprehensive}. Unlike deep neural networks, the key idea behind such generalization is the message passing framework that smooths the message with respect to the local neighborhood \cite{gilmer2017neural}. The design of message passing are majorly motivated in spatial domain \cite{scarselli2008graph,derr2018signed} and 
spectral domain \cite{kipf2016semi,defferrard2016convolutional}. GNNs literature broadly includes convolutional layers \cite{kipf2016semi,liu2021graph}, 
aggregation operators \cite{hamilton2017inductive}, 
pooling methods \cite{zhang2018end}, 
and feature augmentation \cite{chen2020iterative,said2023neurograph,saidgraph2023}. 
There is a growing interest in minimizing the vulnerability of GNNs to node feature and graph structure noise in recent years. A few notable works in this direction include \cite{liu2021graph, zugner2018adversarial, dai2018adversarial}. Similarly, numerous approaches for alleviating structural noise in GNN settings have been proposed, including \cite{zhu2019robust, entezari2020all}. For further reading, please refer to the comprehensive survey of adversarial attacks on graphs \cite{jin2021adversarial,jin2020self}.

Unlike the existing techniques, we pursue a novel approach to advance graph learning in social networks through the incorporation of 
subgraph embeddings, with a primary focus on applications where crucial node features are absent. To this end, we have introduced a learning framework that integrates GNNs 
with graph descriptors, thereby presenting a comprehensive methodology that is adaptable to all types of graph embeddings and GNN architectures. Our evaluations in both node and graph classification settings show encouraging results.


\vspace{-0.5ex}
\section{Preliminaries}
\label{preliminaries}
\vspace{-0.95ex}
Let $G = (V, E, X)$ denote a graph with a set of nodes $V$, edges $E$ and a node feature matrix $X \in \mathbb{R}^{n\times d}$, where $n$ is the total number of nodes and $d$ is the dimension of the feature vector associated with each node. We represent the feature vector associated with the node, $v$, by $x_v$. For some positive integer $k$, let $N_v(k)$ be the set of nodes in the $k$-neighborhood of node $v$, i.e.,  nodes that are at most distance $k$ from $v$. $N_v(k)$ also includes $v$. For a fixed $k$, let $s^k_v$, or $s_v$ when $k$ is clear from context, be an induced subgraph of $G$ on $N_v(k)$. Let $\mathcal{S} = \{s_1,s_2,\cdots,s_n\}$ be the family of all such subgraphs. We refer to $s_v$ as a \emph{$k$-order subgraph} at node $v$. Let $L$ indicates the Laplacian matrix of the graph and $\Phi$ is the matrix consisting of normalized and mutually orthogonal eigenvectors of $L$. Let $\Lambda$ represents the diagonal matrix of the eigenvalues of the Laplacian. We define a subgraph embedding as a function $\phi$ that extracts a compact and expressive signature of a $k$-order subgraph, $\phi:\mathcal{G} \rightarrow \mathbb{R}^{d^\prime}$, where $\mathcal{G}$ is the family of all finite graphs and $d^\prime$ is the required dimension of latent feature space, which may be different from 
$d$. 

\section{Ego-centric Spectral subGraph Embedding Augmentation Framework}\vspace{-2pt}
In this section, we outline the details of our Ego-centric Spectral subGraph Embedding Augmentation (ESGEA) framework. The proposed scheme is divided into 
four phases: (1) ego-centric subgraph extraction around each node, (2) subgraph embedding design $(\phi)$, (3) feature augmentation and, (4) learning module. In the forthcoming sections, we outline the details of these phases one by one. \vspace{-1.0ex}
\subsection{Ego-Centric Subgraph Extraction}
\label{subsec:subgraphs-extraction}
\vspace{-1.0ex}
Locally induced subgraphs capture the topological information of nearby nodes, enabling similar representations for nodes with identical subgraphs. They encode distinct attributes that are not always determined by explicit node features or properties based on the overall topology of the network. Subgraphs feature non-trivial internal structure, border connectivity, and concepts of neighborhood and position in relation to the remainder of the graph. They are, therefore, conducive to learning effective graph representations. 

The importance of subgraphs can be further motivated by the famous reconstructibility conjecture, which holds true for many graph families, including but not limited to regular graphs, trees, Eulerian graphs, outer planner graphs, and graphs with at most 9 vertices~\cite{spinoza2019reconstruction,bilinski2007reconstruction,bondy1977graph}. The conjecture states the following:

\vspace{0.07in}
\noindent
\textbf{Conjecture \cite{kelly1942isometric,ulam1960collection,bondy1977graph}} \emph{A graph with at least three vertices can be constructed uniquely (up to isomorphism) from a collection of its vertex-deleted subgraphs}. 

\vspace{0.07in}
In other words, if $s_v$ is a subgraph of $G=(V,E)$ obtained by deleting vertex $v$ and its incident edges, then $s$ has a unique (up to isomorphism) collection of vertex deleted subgraphs $\{s_1,s_2,\cdots,s_n\}$. Variants of this conjecture deal with different subgraphs of $s$ (e.g.,~\cite{levenshtein2008,levenshtein2008conjecture}). The primary assertion here is that \emph{the topological and structural properties of a graph can ensue from its subgraphs.}

\begin{figure}[!t]
    \centering
    \includegraphics[width = 0.47\textwidth]{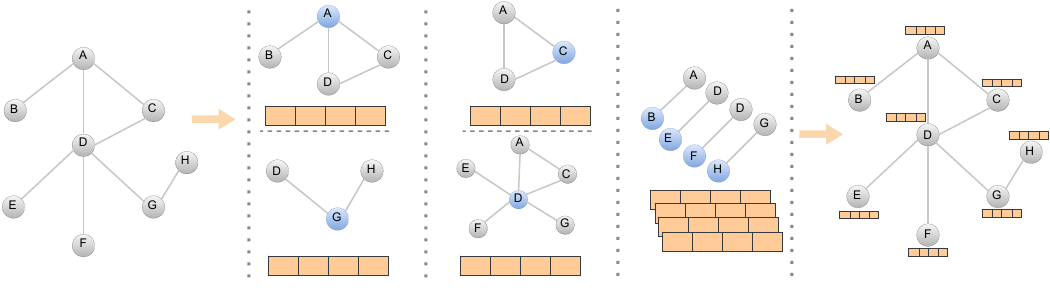}
    \vspace{2.0ex}
    \caption{
    An illustration of constructing the simplest ego-centric subgraph embeddings (with $1$-order subgraphs).
    For each node (indicated in blue), an induced subgraph from its immediate $1$-order neighborhood is extracted and embeddings obtained. The obtained embeddings are finally associated with each node, respectively.  }
    \label{fig:feature-const-example}
    \vspace{6ex}
\end{figure}

This discussion inspires and motivates the study of subgraphs for graph learning. For our purpose, we generate a $k$-order subgraph for each node $v$, which is essentially a subgraph induced on the nodes that are at most distance $k$ from node $v$. Subsequently, we generate an embedding for each node $v$ based on its $k$-order subgraph (as discussed in the next subsection).  
Depending on a node's structural role or position, the corresponding $k$-order subgraph may have different structural features. For instance, as shown in the simplest example $k$=1 in Figure~\ref{fig:feature-const-example}, $1$-order subgraphs of leaf nodes ($B,E,F,$ and $H$) are the same (up to isomorphism). We note that non-leaf nodes generate $1$-order subgraphs distinct from leaf nodes and as $k$ increases more nodes are likely to have distinct embeddings.

This subgraph embedding approach improves the learning of GNNs, particularly when abnormal or no node features are provided.  
Moreover, it enables a fairly broad approach by bridging the gap between GNNs and graph descriptors \cite{said2021dgsd,verma2017hunt,tsitsulin2018netlsd}. 

\subsection{Design of Subgraph Embeddings} 
\label{sec:subgraph-embeddings}
Designing a subgraph embedding that encode structural information at all scales, along with succinctness and expressivity, is a challenging task. Network Laplacian Spectral Descriptor (NetLSD) \cite{tsitsulin2018netlsd} was designed to satisfy most of the required properties that a graph descriptor should possess. For instance, the permutation invariance, extracting small, medium and large scale information, and time and memory efficient. NetLSD is based on the idea of diffusion in graphs, for example, how the heat diffuses across the nodes in the graph. The heat diffusion process over the graph at time $t$ is examined through the \emph{heat kernel} $H_t$, which is the matrix exponential $H_t = e^{-Lt}$, where $L$ is the graph Laplacian and $t$ is the time. Since $L$ can be factorized as, $L = \Phi\Lambda \Phi^\top$, where $\Phi$ is the matrix consisting of normalized and mutually orthogonal eigenvectors, and $\Lambda$ is a diagonal matrix consisting of the eigenvalues of $L$, we obtain the following:
 \vspace{2ex}
\begin{equation}
    \label{eq:heatkernel}
    H_t = e^{-Lt_i} = e^{\Phi {(-\Lambda t_i)}\Phi^\top} = \Phi e^{-\Lambda t_i}\Phi^\top.
\end{equation}
where the $ij^{th}$ entry of $H_t$ indicates the amount of heat transferred from node $v_i$ to $v_j$ at time $t_i$. Similarly, the \textit{wave kernel}, on the other hand, measures the propagation of mechanical waves across the graph. NetLSD is then defined by the traces of the heat or wave kernels at various time intervals: $h_{t_i} = tr(H_{t_i})$, which allows extracting more global information over time. In addition, they are permutation- and size-invariant and scale-adaptive. In terms of time complexity, the Laplacian spectrum requires $O(n^3)$ time and $O(n^2)$ memory, which hinders its scalability.  
Thus, the authors opted for block Krylov-Schur implementation in SLEPc \cite{hernandez2005slepc,said2023augmenting} to compute only $\lambda$ extreme eigenvalues, thereby reducing the computation time and making NetLSD a fitting choice for the subgraph embedding.


\subsection{Feature Aggregation with ESGEA}
\label{sec:feat-aggr}
Aggregation functions play a crucial role in the representation and modeling of graph-structured data, specifically in the message passing framework, and has received significant attention in the literature \cite{hamilton2017inductive}. The performance and representational power of the models are significantly influenced by the selection of aggregation functions. For instance, several studies show that \textit{sum} aggregation allows learning of graph structural properties \cite{xu2018powerful}. Likewise, the \textit{mean} aggregation is often employed to capture the distribution of the elements under consideration, while the \textit{max} aggregation is commonly utilized to identify the most representative elements \cite{xu2018powerful}. Given $x_v$ and $\phi(s_v)$, we define 
a general aggregation function as follow:
\vspace{2ex}
\begin{equation*}
    x_v^\prime = f(x_v, \phi(s_v))
\end{equation*}
$x_v$ and $\phi(s_v)$ correspond to the node feature vector and subgraph embeddings, respectively. The function $f(.)$ can be substituted with aggregation such as \textit{mean, max, min} or \textit{sum}, depending upon the choice of the method. Because simple aggregations like \textit{mean, max} and \textit{sum} may result in the loss of valuable information, several recent studies have suggested using multiple aggregations, feature concatenation \cite{hamilton2017inductive}, aggregation in hierarchical fashion, as well as learnable aggregations \cite{li2020deepergcn}. Learnable aggregations typically entail the utilization of different techniques such as multi-layered perceptron or deep neural networks, which are capable of encoding intricate details and nuances of the input representations. Such methods have demonstrated significant potential in achieving superior performance in a wide range of applications, and are thus the subject of extensive research and investigation in the field. 

\begin{figure}[!t]
    \centering
    \includegraphics[width = 0.47\textwidth]{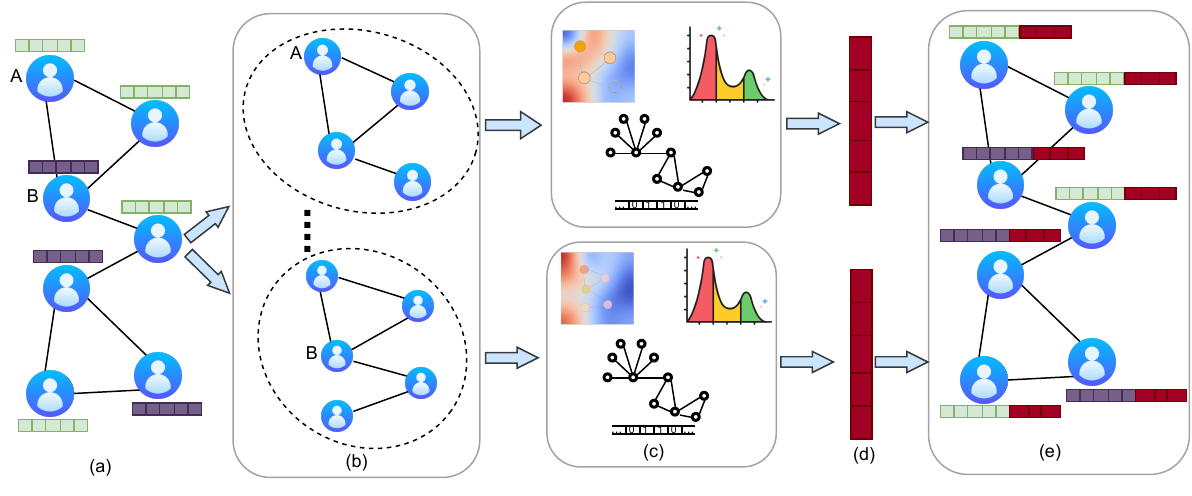}
    \vspace{1ex}
    \caption{Illustration of the overall ego-centric subgraph extraction, embeddings and aggregation: (a) is the input graph with normal (green) and abnormal (gray) node features; (b) demonstrates step 1, i.e., the generation of $k-$order subgraphs for each node; (c) shows the use a set of graph descriptors for extracting graph embeddings; (d) indicates the corresponding embeddings extracted for each subgraph; and (e) illustrates feature aggregation (concatenation in this case) of subgraphs feature vectors with their original features.}  
    \label{fig:architecture-diag}
    \vspace{0.3in}
\end{figure}

In our proposed framework, we offer an adaptable aggregation module that can integrate any of the current aggregation operators to amalgamate the embeddings obtained from the subgraph to the primary node embeddings. However, it is crucial to take into account the limitations of these operators before deciding on the most suitable aggregation approach. While the use of learnable aggregations has its advantages, it is important to note that these methods can be computationally complex, posing challenges in terms of scalability and efficiency. Additionally, the traditional \textit{mean, max}, and \textit{sum} aggregation operators may not always be appropriate, especially in cases where the dimensions of the embeddings ($d^\prime$ and $d$) are unequal. In such cases, a simple combine function such as feature concatenation is a viable alternative, which can work effectively in all scenarios and potentially yield improved results. Figure \ref{fig:architecture-diag} illustrates our methodology for extracting subgraph, computing spectral embeddings, and feature augmentation. The illustration shows a graph with three corrupted (in gray) and four complete node features (in green) and highlights the overall subgraph feature extraction and augmentation approach. 

\subsection{Graph Learning Module for ESGEA}
\label{sec:gnns}
This section focuses on \textit{Message Passing Graph Neural Networks (MPNNs)}, which uses an iterative learning approach to acquire graph representations. MPNNs retain a representation vector $h_v^l \in \mathbf{R}^{\hat{d}}$ for each node $v \in V$ in a given a graph $G = (V, E, X)$. Note that $\hat{d}$ is the size of the embedding $h$, which may vary from $d$ and $d^\prime$ which are the sizes of the input 

\begin{algorithm}[!t]
\DontPrintSemicolon
\footnotesize
\KwIn{Graph $\mathcal{G} = \{g_1, g_2, \dots, g_n\}$, subgraph depth: $k$,  \#Layers: $L$,  non-linearity: $\sigma$  Weight matrices $W^l$ }
\KwOut{vector output $z_g$ for all $g \in \mathcal{G}$}
\SetKwBlock{Begin}{function}{end function}
\For{$G$ in $\mathcal{G}$}
{
    \For{node $v$ in $G$}
    {
      $s_v \leftarrow Extract\_subgraph(G,v,k)$;\\
      $h_v^\prime \leftarrow \phi(s_v)$\\
      $x_v^\prime \leftarrow h_v^\prime$ \\
    }\label{endfor}
    $h_v^{(0)} \leftarrow x_v^\prime \, \,\forall \, v \in G$\\
}\label{endfor}

\For{$G$ in $\mathcal{G}$}
{
    \For{$l =1, \ldots, L$}
    {
        \For{node $v$ in $G$}
        {
            $a_v^{(l)} = f_\text{AGG} \left(h_u^{(l-1)} \, | \, u \in \mathcal{N}(v)\right) $ \\
            $h_v^{(l)} = f_{\textit{UPDATE}}^{(l)}\left( h_v^{(l-1)},a_v^{(l)}\right)$\\
        }\label{endfor}
    $h_g^{(l)} \leftarrow h_v^{L} \, \forall \, v \in G$    
    }\label{endfor}
}\label{endfor}
  \Return{$z_g \leftarrow h_g \, \forall \, g \in \mathcal{G}$ }
\caption{ESGEA for Graph Classification}
\label{algo:graph-classification}

\end{algorithm}
\vspace*{1ex}

\noindent features and subgraph embeddings respectively. MPNNs are defined as follows: 
\begin{align}
h_v^{(0)} = x_v \, \forall v \in V \\
a_v^{(l)} = f_{\textit{AGG}}^{(l)}\left(h_u^{(l-1)} | u \in \mathcal{N}(v)\right) \\
h_v^{(l)} = f_{\textit{UPDATE}}^{(l)}\left( h_v^{(l-1)},a_v^{(l)}\right)
\end{align}
    
Node features $h_v^{(0)}$ are initialized with the original node features $x_v$ and then the aggregate and update functions are used to update the node features based on its neighbors, and the prior state at every iteration. 

In Algorithm \ref{algo:graph-classification}, we provide a step by step procedure of the proposed framework in a graph classification setting. The algorithm first extracts subgraph embeddings for each node using an input parameter $k$, which is the depth of the subgraphs. These subgraphs are then supplied to the embedding function to produce graph embeddings. We would like to note that we propose spectral graph embeddings, e.g., NetLSD \cite{tsitsulin2018netlsd} for generating subgraph embeddings which distinguishes our work from the existing nested subgraphs GNN methods. As described in section \ref{sec:subgraph-embeddings}, spectral descriptors are powerful methods for extracting expressive graph embeddings. Nonetheless, our proposed framework is general, thus any embedding method may be used in this step. As we do not consider node features in social networks, we assign subgraph embeddings $h_v^\prime$ as node features in the graph classification setting (step 5 and 8). In addition, we provide a generalized MPNNs strategy for learning representations for each graph that may be used for the subsequent ML task. $f_{AGG}(.)$ and $f_{UPDATE}(.)$ are generic functions that may be fine-tuned based on the learning architecture of choice.

Similar to the graph classification, node classification is a well-studied problem, particularly in semi-supervised learning. It has numerous applications in several fields, such as online social networks, biological networks, and ecommerce networks. Node classification involves training a model in a supervised or semi-supervised setting that can predict a label for a new unseen node. In a GNN setting, we obtain node representations from the last layer followed by a linear layer to obtain the class label. A loss function is then applied to train the model accordingly. We define the cross entropy loss function as follows that we use for binary classification. 
\vspace{2ex}
\begin{equation}
    \mathcal{L} = -\frac{1}{N}\sum_{j=v}^N y_j \, log (p_j) +  (1-y_j)\,log(1-p_j)
\end{equation}
where $p_i$ is the Softmax probability obtained for the data point $j$, and $y_j$ is the corresponding ground truth value. 

To adapt Algorithm \ref{algo:graph-classification} to the node classification task, a few modifications can be made. Specifically, the subgraph embeddings generated by the embedding function (step 4) are integrated with the node features using an aggregation function, which is thoroughly explained in \ref{sec:feat-aggr}. The resulting aggregated node features are then fed into the learning framework to obtain node representations. In contrast to the graph classification setting, where we process a collection of graphs (step 1 and 7), in this case we iterate solely over the nodes of a single graph and learn node embeddings. Thus, the proposed learning framework provides a flexible MPNNs-based solution for node classification as well.

\begin{figure}
    \centering
    \includegraphics[width =0.49\textwidth]{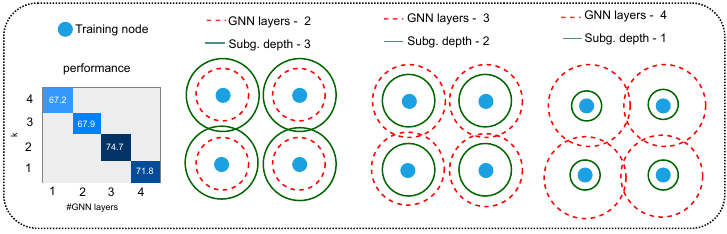}
    \caption{An illustration of the comparative merits of GNNs and subgraph depth with respect to their receptive fields within the graph and resultant performance reported on LastFM Asia dataset.}
    \label{fig:subgnn-illustration}
    \vskip 4ex
\end{figure}

The Algorithm \ref{algo:graph-classification} introduces a novel framework that primarily involves two crucial input parameters impacting the model's performance: (a) the depth of subgraphs $(k)$ and the number of GNN layers $L$. The combination of these parameters plays a crucial role in feature aggregation and the receptive fields of the models. Figure \ref{fig:subgnn-illustration} demonstrates the comparative merits of different combinations of these parameters. A larger value of $k$ and $L$ for a given training node can increase the overlap of information obtained from the node's neighborhood. In addition, it increases the model's receptive field and may result in oversmoothing that hinders the model's performance. Similarly, a combination of large and small input values widens the distance between the embeddings and message passing, which may result in a reduction in performance. 
Conversely, combining both parameters can result in superior performance.

\vspace{-1ex}
\section{Experimental Evaluation}
\label{sec:evaluation}
\vspace{-1ex}

\begin{figure*}[!t]
    \hspace*{-1.75ex}\includegraphics[width = 1.025\textwidth]{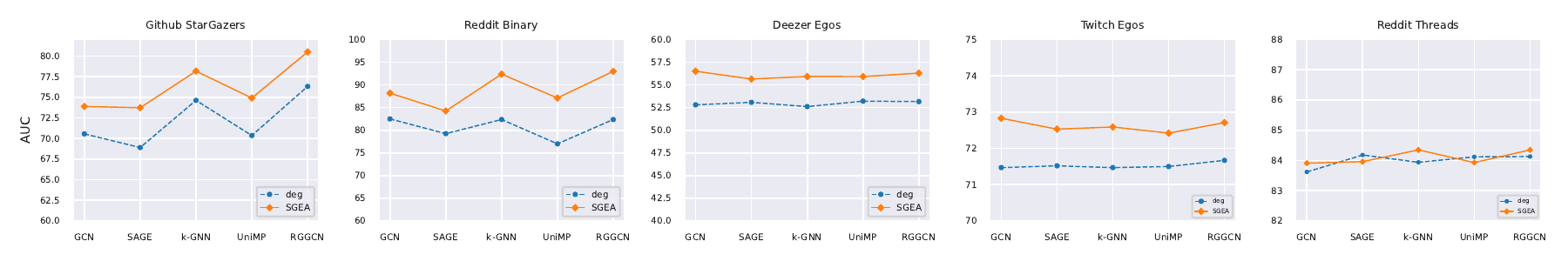}
    \caption{Comparison of mean AUC scores (10 runs with different random seeds) against \emph{one-hot-degree-encoding} ($deg$) as node features on five different GNN models in graph classification setting. }
    \label{fig:GC_plots}
    \vspace{2.5ex}
\end{figure*}
To assess the performance of the proposed framework, here we define the following two research questions. 
\begin{itemize}
 \item \textbf{RQ-1} Can the proposed ESGEA framework improve performance in the absence of explicit node features?

 \item \textbf{RQ-2} Does the effectiveness of the proposed ESGEA framework remain intact to improve performance in applications where node features are unreliable?
\end{itemize}

To answer the first question, we consider a graph classification setting in which node features are unavailable. For the second research question, we consider node classification setting with abnormal node features. The forthcoming sections detail the experimental design and results. 

\textbf{General Setup:} 
We ran all the experiments on a $112$-core Intel Xeon CPU 2.20 GHz machine with 512 GB of RAM and an Nvidia GPU with 48 GB of memory. Each method is trained on $80\%$ and tested on $20\%$ of the dataset. 
We run all methods with 10 different seeds, and average accuracy and Area Under the Curve (AUC) are reported for node classification and graph classification tasks, respectively. Throughout the experiments, we consider the depth of the subgraphs to be $3$ and use NetLSD descriptor to construct subgraph embeddings. The source code is made publicly available\footnote{ESGEA publicly available code: \url{https://github.com/Anwar-Said/ESGEA}}.

\subsection{Graph Classification}
\label{subsec:graph-classification}
When node features are unavailable, current graph classification techniques typically rely on one-hot-degree encoding to provide relevant node features for GNNs in order to improve their performance. However, in the majority of cases, these features 
do not provide a stronger learning basis for GNNs. Here, we hypothesize that the features generated through the proposed approach improve the performance of models for datasets without node features. We test our hypothesis in the graph classification setting on five datasets where node features are unavailable. 

\textbf{Datasets:} We consider \textit{Github Stargazers}, \textit{Reddit threads}, \textit{Reddit Binary}, \textit{Deezer Egos}, and \textit{Twitch Egos} social network datasets in our experimental setup. 
Due to space limitation, we refer the reader to \cite{karateclub} for 
further dataset details. 



\textbf{Baselines:} 
We consider the following backbone GNN models, including which include the seminal GCN, GraphSAGE, and Residual Gated GCN (RGGCN), along with more recent advanced model UniMP and provably more expressive 
k-GNN.

\textbf{Experimental setup:} Initially we generate node features for each node with subgraphs embedding using NetLSD descriptor. The dimension of NetLSD descriptor is set to $20$ and depth of the subgraphs $k$ is set to $2$ (small diameters) and $3$. We then run each of these models with the generated NetLSD embeddings and one-hot-degree encoding and compare the results. Each model consists of three convolution layers followed by SortPooling layers \cite{zhang2018end}, two 1D convolutions and three MLP layers. The train:test splits ratio was set to $80:20$, batch size to $128$, learning rate to $1e^{-4}$ and the number of epochs was set to $100$. We consider Area Under the Curve (AUC) similar to \cite{karateclub} as the evaluation metric.

We report the performance comparison in terms of AUC with degree encoding as node features in Figure \ref{fig:GC_plots}. The results demonstrate that ESGEA vastly outperforms one-hot-degree encoding. More specifically, Residual Gated GCN and $k$-GNN demonstrate an encouraging improvement of up to $10\%$ on the Reddit binary dataset. Similarly, each method has received up to $4\%$ improvement on the Github StarGazers dataset. ESGEA also outperforms throughout on Twitch Egos and Deezer Egos datasets. These results clearly illustrate the effectiveness of the proposed approach in the applications where node features are unavailable.

\begin{table}[!tb]
\centering
\footnotesize
\vspace{0.3cm}
\caption{Comparison of node classification accuracy on the test set of Facebook dataset across 10 seeded runs\protect\footnotemark[1].} 
\vspace{1.5ex}
\begin{tabular}{|l|l|l|l|l|l|l|}
\hline
&\multicolumn{6}{c|}{\textbf{Corruption ratio}} \\ \cline{2-7}
\multicolumn{1}{|c|}{\textbf{Model}} & \multicolumn{1}{c|}{\textbf{0\%}} & \multicolumn{1}{c|}{\textbf{10\%}} & \multicolumn{1}{c|}{\textbf{20\%}} & \multicolumn{1}{c|}{\textbf{50\%}} & \multicolumn{1}{c|}{\textbf{80\%}} & \textbf{90\%} \\ \cline{1-7}
MLP & 76.30 & 67.96 & 61.48 & 43.29 & {\color{blue}32.84} & {\color{blue}30.70} \\ 
ESGEA & {\color{red}76.81} & {\color{red}68.49} & {\color{red}62.10} & {\color{red}43.51} & 31.14 & 30.56 \\ \cline{1-7} 
GCN & 93.88 & 92.01 & 89.89 & 83.60 & 77.08 & 75.60 \\ 
ESGEA & {\color{red}93.89} & {\color{red}92.08} & {\color{red}90.16} & {\color{red}83.67} & {\color{red}78.23} & {\color{red}76.75} \\ \cline{1-7} 
GAT & 93.86 & 85.06 & 72.80 & 59.01 & 54.85 & 59.82 \\
ESGEA & {\color{red}94.01} & {\color{red}85.96} & {\color{red}81.09} & {\color{red}67.77} & {\color{red}66.19} & {\color{red}63.03} \\ \cline{1-7} 
AirGNN & 90.01 & {\color{blue}88.06} & {\color{blue}86.27} & 79.53 & 64.80 & {\color{blue}61.46} \\ 
ESGEA & {\color{red}90.02} & 87.80 & 86.14 & {\color{red}79.54} & {\color{red}66.15} & 58.61 \\  \cline{1-7} 
UniMP & 95.16 & {\color{blue}94.44} & 93.05 & 85.03 & 78.44 & 77.80 \\ 
ESGEA & {\color{red}95.27} & 94.31 & {\color{red}93.22} & {\color{red}89.41} & {\color{red}85.89} & {\color{red}85.13} \\ \hline
\end{tabular}%
\vspace{3.0ex}
\label{res:facebook-results}
\end{table}

\footnotetext[1]{The results are colored {\color{red}red} where the proposed method ESGEA outperforms the baseline and are colored {\color{blue}blue} otherwise}

\vspace{-1ex}
\subsection{Node Classification}
\label{subsec:node-classification}
In the node classification setting, we evaluate the proposed method in a setting where a certain percentage of node features is corrupted. We apply the proposed approach to enrich the corrupted node features and then trained different GNNs to evaluate the results. In the following sections, we describe the datasets, experimental setup, and results in detail.

\begin{table}[!ttb]
 \vspace{2.5ex}
\centering
\footnotesize
\caption{Comparison of node classification accuracy on the test set of LastFM Asia dataset across 10 seeded runs\protect\footnotemark[1].}
\vspace{1.5ex}
\begin{tabular}{|l|l|l|l|l|l|l|}
\hline
&\multicolumn{6}{c|}{\textbf{Corruption ratio}} \\ \cline{2-7}
\multicolumn{1}{|c|}{\textbf{Model}} & \multicolumn{1}{c|}{\textbf{0\%}} & \multicolumn{1}{c|}{\textbf{10\%}} & \multicolumn{1}{c|}{\textbf{20\%}} & \multicolumn{1}{c|}{\textbf{50\%}} & \multicolumn{1}{c|}{\textbf{80\%}} & \textbf{90\%} \\ \cline{1-7}
MLP & 71.76 & 59.50 & 49.43 & {\color{blue}31.13} & {\color{blue}21.62} & 20.24 \\ 
ESGEA & {\color{red}71.85}& {\color{red}60.00} & {\color{red}50.39} & 30.32 & 20.98& {\color{red}20.25} \\ \cline{1-7} 
GCN & 86.11 & 84.28 & 82.36 & 74.94 & 67.50 & 66.03 \\ 
ESGEA & {\color{red}86.50} & {\color{red}84.55} & {\color{red}86.47} & {\color{red}75.43} & {\color{red}68.04} & {\color{red}66.89} \\ \cline{1-7} 
GAT & 85.52 & 79.37 & 78.03& 72.20 & 67.65 & 70.91 \\ 
ESGEA & {\color{red}85.72} & {\color{red}79.52} & {\color{red}85.72} & {\color{red}75.42} & {\color{red}73.21} & {\color{red}75.24} \\ \cline{1-7} 
AirGNN & {\color{blue}86.45} & {\color{blue}85.76} & 85.69 & 82.31 & {\color{blue}79.37} & 77.17 \\ 
ESGEA & 86.39 & 85.70 & {\color{red}85.70} & {\color{red}83.35} & 78.50 & {\color{red}77.36} \\  \cline{1-7} 
UniMP & 87.12 & 85.69 & 82.98 & 77.47 & 72.07&71.70 \\ 
ESGEA & {\color{red}87.28} & {\color{red}84.58} & {\color{red}87.20} & {\color{red}80.61} & {\color{red}78.03} & {\color{red}78.54} \\ \hline
\end{tabular}%
\label{tab:lastfmasi-res}
\end{table}
\begin{figure}[!t]
    \centering
    \hspace{-1.75ex}\includegraphics[width = 0.5\textwidth]{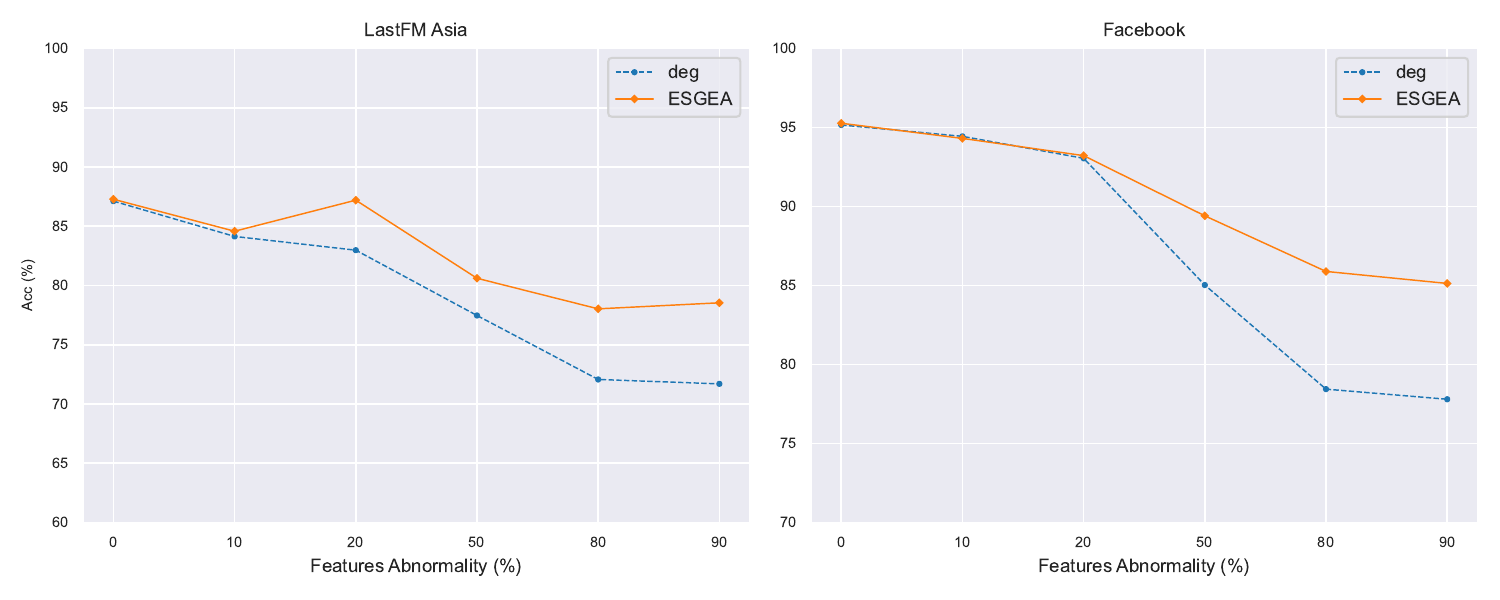}
    \vspace{0.1ex}
    \caption{ESGEA with UniMP performance on both Facebook and LastFM Asia datasets. Here we vary 
    the percent of abnormality injected into the original node features. The results are compared with the standard one-hot-degree encoding (denoted as deg). }
    \label{fig:unimp_plot}
    \vskip 5ex
\end{figure}

\textbf{Datasets:} We consider two social network datasets, \textit{Facebook Page-page} and \textit{LastFM Asia}, 
for the node classification task. We refer the reader to \cite{rozemberczki2021multi} for further details on these datasets. 

\textbf{Backbone GNNs:} 
In our evaluations, we consider four baseline methods: a two-layered MLP, GCN \cite{kipf2016semi}, GAT \cite{velivckovic2017graph}, Unified Message Passing (UniMP) \cite{shi2020masked}, and AirGNN \cite{liu2021graph}. 



\textbf{Setup:}  
We examine a three-layered architecture for every method implemented in PyTorch Geometric. The number of hidden channels was set to 16,  learning rate as $0.01$, $k$ and $L$ equal $3$ and the weight decay was set to $1e^{-4}$. Each model was trained for $200$ epochs with $50$ epochs as an early stopping criterion. Throughout the experiments, we evaluate all models on $0\%, 10\%, 20\%, 50\%, 80\%$, and $90\%$ noise setting, i.e., the number represents the percentage of nodes for which original node features were replaced with random features sampled from a Gaussian distribution as done in \cite{liu2021graph}.  

\textbf{Results:} Table \ref{res:facebook-results} and \ref{tab:lastfmasi-res} present the classification results of our evaluation on both datasets. The proposed framework achieves either comparable or superior performance in every experiment. 
Specifically, the proposed method with UniMP and GAT models boasts a performance gain of up to $7\%$ on the Facebook and LastFM Asia datasets. 
In Figure \ref{fig:unimp_plot}, we visualize the performance of ESGEA in conjunction with 
UniMP on both Facebook and LastFM Asia datasets. We observe as the abnormality increases, ESGEA is less impacted as degree encoding and maintains stronger performance.

\subsection{Parameters Sensitivity and Runtime}
\label{sec:parameters-sensitivity}

\begin{figure}[!b]
\centering
\begin{minipage}{.258\textwidth}
  \hspace{-2.0ex}\includegraphics[width=1.0\textwidth]{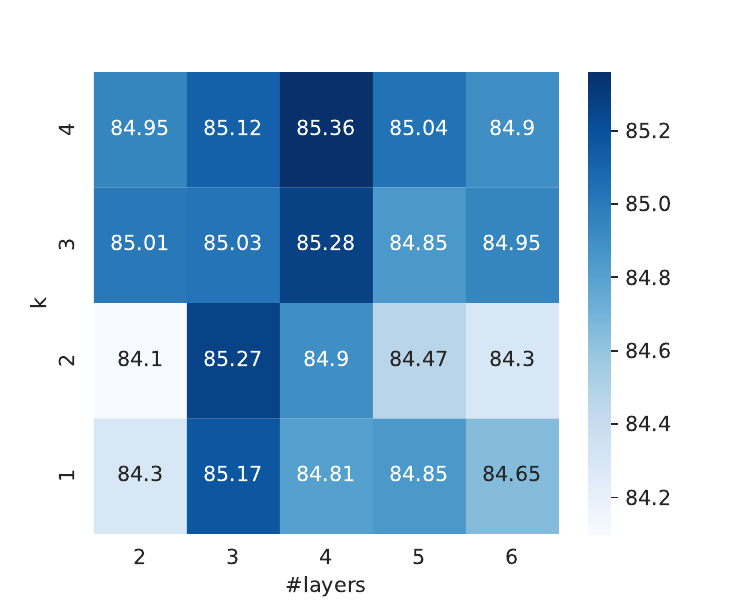}
\end{minipage}%
\begin{minipage}[width=0.235\textwidth]{.268\textwidth}
  \centering
  \hspace{-9ex}\includegraphics[width=1.0\textwidth]{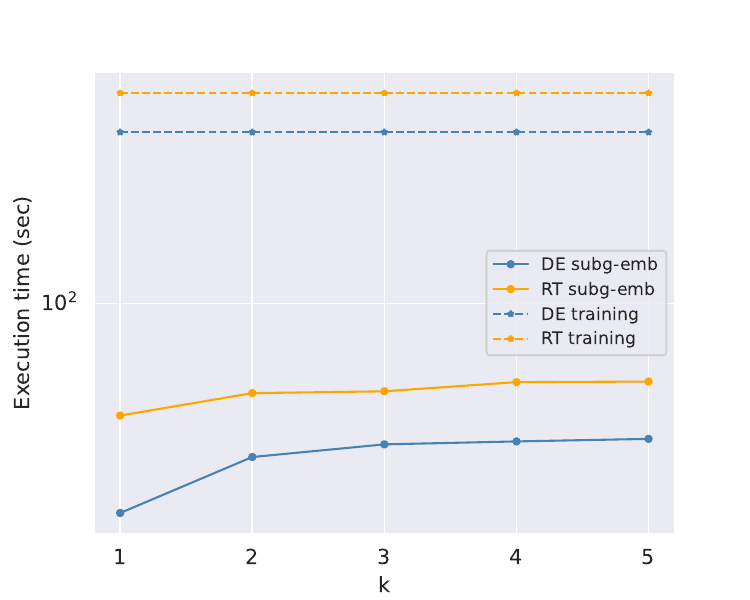}
\end{minipage}
 \vspace{0.1in}
\caption{Graph classification parameter sensitivity and runtime analysis: (a)  Performance comparison of $k-$GNN on Reddit threads dataset. (b) running times of subgraph embeddings and training times of GCN on Reddit threads and Deezer egos datasets. }
\label{fig:sensitivity-analysis-gc}
\vspace{0.3in}
\end{figure}
\vspace{-0.07in}
Graph neural networks that employ subgraph representations are a novel and sophisticated category of expressive learning techniques designed to represent graphs as an amalgamation of subgraphs \cite{frasca2022understanding}. The efficacy of these methods is typically contingent upon the depth of the subgraphs as well as the number of GNN layers utilized, both of which constitute hyper-parameters. As the number of layers in MPNNs increases, a common phenomenon known as oversmoothing occurs, whereby the node features begin to converge into indistinguishable vectors. This is evidenced in Figure \ref{fig:examples}, which shows the performance of GCN on the Facebook dataset. Conversely, while a deeper subgraph results in an enlarged receptive field, it also exacerbates the oversmoothing effect, as node features are smoothed too quickly. Moreover, it also increases the running times of the methods by several folds. It is therefore imperative to make informed selections for these hyper-parameters in order to facilitate the training of the model. To delve deeper into the interplay of these parameters, we conducted an empirical analysis on both the node and graph classification tasks. 

\textbf{Graph Classification:} We use Reddit threads dataset to analyze $k-$GNN with varying numbers of layers and subgraph depths, as illustrated in Figure \ref{fig:sensitivity-analysis-gc} (a). Our objective was to determine how these parameters affect performance. We observe that a smaller or larger number of GNN layers reduces the model's performance, whereas a layer count of $3$ or $4$ yields optimal results. We also assessed the execution time of subgraph embedding and model training on the Deezer egos and Reddit threads datasets, as displayed in Figure \ref{fig:sensitivity-analysis-gc} (b). These datasets have small diameters averaging at $3.4$ and $4.5$. As a consequence of their low diameters, the computation time for subgraph embedding is not extensive and can be completed within $50$ seconds. Similarly, model training times fall within the range of $200-300$ seconds. These findings demonstrate that computing embeddings using subgraphs and NetLSD descriptors is scalable on large graphs, without sacrificing the method's efficiency.

\begin{figure}[!t]
\centering
\vspace{-2.5ex}
\begin{minipage}{.268\textwidth}
  \hspace{-2.0ex}\includegraphics[width=1.0\textwidth]{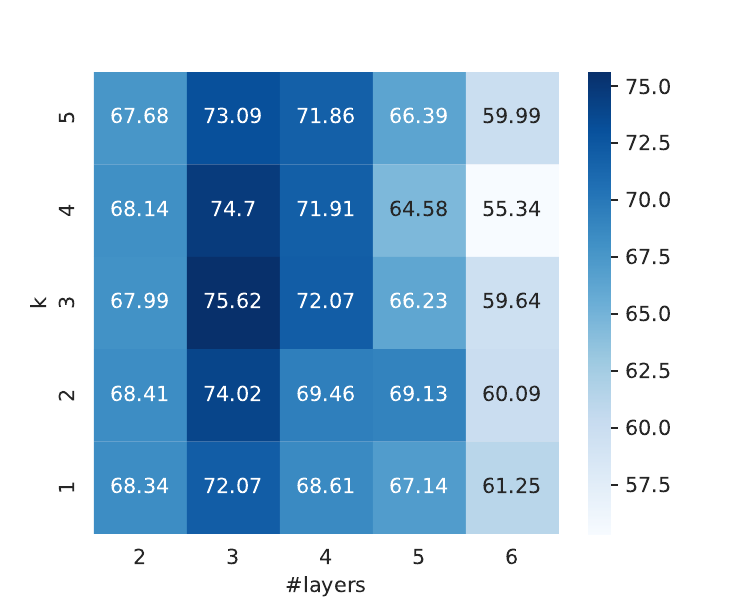}
\end{minipage}%
\begin{minipage}[width=0.235\textwidth]{.268\textwidth}
  \centering
  \hspace{-9ex}\includegraphics[width=1.0\textwidth]{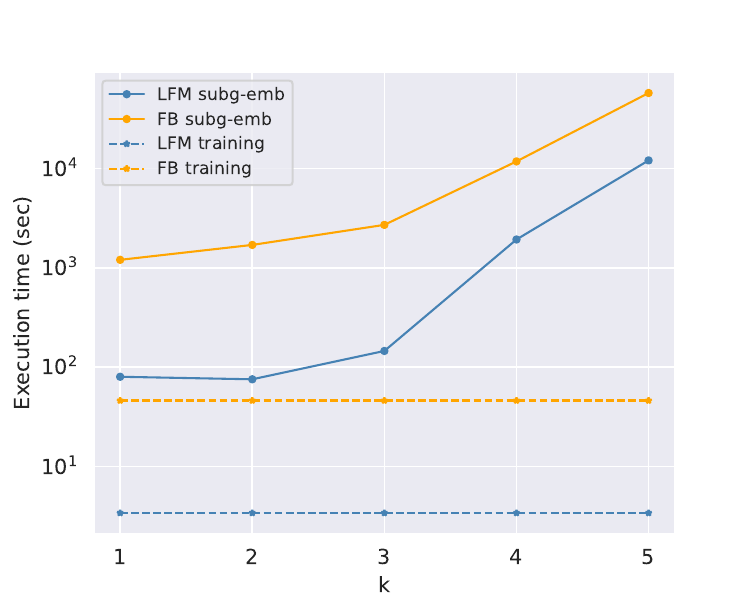}
\end{minipage}
 \vspace{0.125in}
\caption{Node classification parameter sensitivity and runtime analysis: (a)  Performance comparison of Graph attention network on LastFM Asia dataset. (b) running times of subgraph embeddings and training times of GCN on LastFM Asia and Facebook datasets. }
\label{fig:sensitivity-analysis-nc}
\vspace{0.45in}
\end{figure}

\textbf{Node Classification:}
To assess the effectiveness of the hyper-parameters and analyze the running times, we consider both Last FM Asia and Facebook social networks in a node classification setting. As illustrated in Figure \ref{fig:sensitivity-analysis-nc} (a), our results indicate increasing both the number of layers and the depth of subgraphs can lead to suboptimal performance. Our analysis suggests a favorable combination of hyper-parameters involves three GNN layers with depth $k = 3$ for optimal results. In Figure \ref{fig:sensitivity-analysis-nc} (b), we present the running time of the NetLSD descriptor for computing subgraphs of varying depth on the LastFM Asia and Facebook datasets. We also compare the training times of GCN on both datasets.  Notably, as the depth of the subgraphs increases, their size grows, which consequently extends the running time of the descriptor as depicted in Figure~\ref{fig:sensitivity-analysis-nc} (b). Nonetheless, we observe that subgraph embeddings for $k=3$ were calculated within a reasonable $2$ and $45$ minutes accordingly.

\vspace{-0.25ex}
\section{Conclusion}
\vspace{-0.25ex}
In this paper, we proposed Ego-centric Spectral subGraph Embedding Augmentation (ESGEA), a novel framework for extracting node features from network topology, which is especially important for settings where networks have missing or unreliable node feature information. Preceding message passing, our framework leverages the node's neighborhood topological structure from spectral graph embeddings obtained on extracted ego-centric $k$-order subgraphs to generate informative and expressive node features, which are then augmented to the feature matrix using a suitable aggregation approach. Notably, our framework is flexible and compatible with any GNN-based architecture, and exhibits impressive performance. Our detailed evaluations on seven datasets, and eight baselines, encompassing both node and graph classification settings, illustrate the efficacy and potential of the proposed approach. 

\vspace{-0.25ex}
\section*{Acknowledgment}
\vspace{-0.25ex}
This material is based upon work supported by the National Science Foundation Grant Nos. 2239881, 2325416, and 2325417. 

\bibliographystyle{plain}
\bibliography{references}

\begin{thebibliography}{10}

\bibitem{bilinski2007reconstruction}
Mark Bilinski, Young~Soo Kwon, and Xingxing Yu.
\newblock On the reconstruction of planar graphs.
\newblock {\em Journal of Combinatorial Theory, Series B}, 2007.

\bibitem{bondy1977graph}
J~Adrian Bondy and Robert~L Hemminger.
\newblock Graph reconstruction—a survey.
\newblock {\em Journal of Graph Theory}, 1(3):227--268, 1977.

\bibitem{bronstein2017geometric}
Michael~M Bronstein, Joan Bruna, Yann LeCun, Arthur Szlam, and Pierre
  Vandergheynst.
\newblock Geometric deep learning: going beyond euclidean data.
\newblock {\em IEEE Signal Processing Magazine}, 34(4):18--42, 2017.

\bibitem{cappart2021combinatorial}
Quentin Cappart, Didier Ch{\'e}telat, Elias Khalil, Andrea Lodi, Christopher
  Morris, and Petar Veli{\v{c}}kovi{\'c}.
\newblock Combinatorial optimization and reasoning with graph neural networks.
\newblock {\em arXiv preprint arXiv:2102.09544}, 2021.

\bibitem{chen2020iterative}
Yu~Chen, Lingfei Wu, and Mohammed Zaki.
\newblock Iterative deep graph learning for graph neural networks: Better and
  robust node embeddings.
\newblock {\em NeurIPS}, 33, 2020.

\bibitem{choma2018graph}
Nicholas Choma, Federico Monti, Lisa Gerhardt, Tomasz Palczewski, Zahra
  Ronaghi, Prabhat Prabhat, Wahid Bhimji, Michael~M Bronstein, Spencer~R Klein,
  and Joan Bruna.
\newblock Graph neural networks for icecube signal classification.
\newblock In {\em 2018 17th IEEE International Conference on Machine Learning
  and Applications (ICMLA)}, pages 386--391. IEEE, 2018.

\bibitem{dai2018adversarial}
Hanjun Dai, Hui Li, Tian Tian, Xin Huang, Lin Wang, Jun Zhu, and Le~Song.
\newblock Adversarial attack on graph structured data.
\newblock In {\em International conference on machine learning}, pages
  1115--1124. PMLR, 2018.

\bibitem{defferrard2016convolutional}
Micha{\"e}l Defferrard, Xavier Bresson, and Pierre Vandergheynst.
\newblock Convolutional neural networks on graphs with fast localized spectral
  filtering.
\newblock {\em Advances in neural information processing systems}, 29, 2016.

\bibitem{derr2020epidemic}
Tyler Derr, Yao Ma, Wenqi Fan, Xiaorui Liu, Charu Aggarwal, and Jiliang Tang.
\newblock Epidemic graph convolutional network.
\newblock In {\em Proceedings of the 13th International Conference on Web
  Search and Data Mining}, pages 160--168, 2020.

\bibitem{derr2018signed}
Tyler Derr, Yao Ma, and Jiliang Tang.
\newblock Signed graph convolutional networks.
\newblock In {\em 2018 IEEE International Conference on Data Mining (ICDM)},
  pages 929--934. IEEE, 2018.

\bibitem{derrow2021eta}
Austin Derrow-Pinion, Jennifer She, David Wong, Oliver Lange, Todd Hester, Luis
  Perez, Marc Nunkesser, Seongjae Lee, Xueying Guo, Brett Wiltshire, et~al.
\newblock Eta prediction with graph neural networks in google maps.
\newblock In {\em Proceedings of the 30th CIKM}, 2021.

\bibitem{entezari2020all}
Negin Entezari, Saba~A Al-Sayouri, Amirali Darvishzadeh, and Evangelos~E
  Papalexakis.
\newblock All you need is low (rank) defending against adversarial attacks on
  graphs.
\newblock In {\em Proceedings of the 13th International Conference on Web
  Search and Data Mining}, pages 169--177, 2020.

\bibitem{frasca2022understanding}
Fabrizio Frasca, Beatrice Bevilacqua, Michael Bronstein, and Haggai Maron.
\newblock Understanding and extending subgraph gnns by rethinking their
  symmetries.
\newblock {\em Advances in Neural Information Processing Systems},
  35:31376--31390, 2022.

\bibitem{gilmer2017neural}
Justin Gilmer, Samuel~S Schoenholz, Patrick~F Riley, Oriol Vinyals, and
  George~E Dahl.
\newblock Neural message passing for quantum chemistry.
\newblock In {\em International conference on machine learning}. PMLR, 2017.

\bibitem{hamilton2017inductive}
Will Hamilton, Zhitao Ying, and Jure Leskovec.
\newblock Inductive representation learning on large graphs.
\newblock {\em Advances in neural information processing systems}, 30, 2017.

\bibitem{hernandez2005slepc}
Vicente Hernandez, Jose~E Roman, and Vicente Vidal.
\newblock Slepc: A scalable and flexible toolkit for the solution of eigenvalue
  problems.
\newblock {\em (TOMS)}, 31(3):351--362, 2005.

\bibitem{jin2020self}
Wei Jin, Tyler Derr, Haochen Liu, Yiqi Wang, Suhang Wang, Zitao Liu, and
  Jiliang Tang.
\newblock Self-supervised learning on graphs: Deep insights and new direction.
\newblock {\em arXiv preprint arXiv:2006.10141}, 2020.

\bibitem{jin2021adversarial}
Wei Jin, Yaxing Li, Han Xu, Yiqi Wang, Shuiwang Ji, Charu Aggarwal, and Jiliang
  Tang.
\newblock Adversarial attacks and defenses on graphs.
\newblock {\em ACM SIGKDD Explorations Newsletter}, 22(2):19--34, 2021.

\bibitem{kelly1942isometric}
Paul~Joseph Kelly.
\newblock {\em On isometric transformations}.
\newblock PhD thesis, University of Wisconsin, 1942.

\bibitem{kipf2016semi}
Thomas~N Kipf and Max Welling.
\newblock Semi-supervised classification with graph convolutional networks.
\newblock {\em arXiv preprint arXiv:1609.02907}, 2016.

\bibitem{kostochka2020reconstruction}
Alexandr~V Kostochka and Douglas~B West.
\newblock On reconstruction of graphs from the multiset of subgraphs obtained
  by deleting $\ell$ vertices.
\newblock {\em IEEE Transactions on Information Theory}, 2020.

\bibitem{levenshtein2008}
Vladimir Levenshtein, Elena Konstantinova, Eugene Konstantinov, and Sergey
  Molodtsov.
\newblock Reconstruction of a graph from 2-vicinities of its vertices.
\newblock {\em Discrete Applied Mathematics}, 2008.

\bibitem{levenshtein2008conjecture}
Vladimir~I Levenshtein.
\newblock A conjecture on the reconstruction of graphs from metric balls of
  their vertices.
\newblock {\em Discrete Mathematics}, 2008.

\bibitem{li2020deepergcn}
Guohao Li, Chenxin Xiong, Ali Thabet, and Bernard Ghanem.
\newblock Deepergcn: All you need to train deeper gcns.
\newblock {\em arXiv arXiv:2006.07739}, 2020.

\bibitem{liu2021graph}
Xiaorui Liu, Jiayuan Ding, Wei Jin, Han Xu, Yao Ma, Zitao Liu, and Jiliang
  Tang.
\newblock Graph neural networks with adaptive residual.
\newblock {\em NeurIPS}, 34:9720--9733, 2021.

\bibitem{rossi2021unreasonable}
Emanuele Rossi, Henry Kenlay, Maria~I Gorinova, Benjamin~Paul Chamberlain,
  Xiaowen Dong, and Michael Bronstein.
\newblock On the unreasonable effectiveness of feature propagation in learning
  on graphs with missing node features.
\newblock {\em arXiv preprint arXiv:2111.12128}, 2021.

\bibitem{rozemberczki2021multi}
Benedek Rozemberczki, Carl Allen, and Rik Sarkar.
\newblock Multi-scale attributed node embedding.
\newblock {\em Journal of Complex Networks}, 9(2):cnab014, 2021.

\bibitem{karateclub}
Benedek Rozemberczki, Oliver Kiss, and Rik Sarkar.
\newblock {Karate Club: An API Oriented Open-source Python Framework for
  Unsupervised Learning on Graphs}.
\newblock In {\em Proceedings of the 29th ACM (CIKM '20)}. ACM, 2020.

\bibitem{said2023neurograph}
Anwar Said, Roza~G Bayrak, Tyler Derr, Mudassir Shabbir, Daniel Moyer, Catie
  Chang, and Xenofon Koutsoukos.
\newblock Neurograph: Benchmarks for graph machine learning in brain
  connectomics.
\newblock {\em arXiv preprint arXiv:2306.06202}, 2023.

\bibitem{saidgraph2023}
Anwar Said, Tyler Derr, Mudassir Shabbir, Waseem Abbas, and Xenofon Koutsoukos.
\newblock Graph unlearning: A review.
\newblock {\em arXiv preprint arXiv:2310.02164}, 2023.

\bibitem{said2021netki}
Anwar Said, Saeed-Ul Hassan, Waseem Abbas, and Mudassir Shabbir.
\newblock Netki: a kirchhoff index based statistical graph embedding in nearly
  linear time.
\newblock {\em Neurocomputing}, 433:108--118, 2021.

\bibitem{said2021dgsd}
Anwar Said, Saeed-Ul Hassan, Suppawong Tuarob, Raheel Nawaz, and Mudassir
  Shabbir.
\newblock Dgsd: Distributed graph representation via graph statistical
  properties.
\newblock {\em Future Generation Computer Systems}, 2021.

\bibitem{said2023augmenting}
Anwar Said, Mudassir Shabbir, Saeed-Ul Hassan, Zohair~Raza Hassan, Ammar Ahmed,
  and Xenofon Koutsoukos.
\newblock On augmenting topological graph representations for attributed
  graphs.
\newblock {\em Applied Soft Computing}, 136:110104, 2023.

\bibitem{scarselli2008graph}
Franco Scarselli, Marco Gori, Ah~Chung Tsoi, Markus Hagenbuchner, and Gabriele
  Monfardini.
\newblock The graph neural network model.
\newblock {\em IEEE transactions on neural networks}, 20(1):61--80, 2008.

\bibitem{shi2020masked}
Yunsheng Shi, Zhengjie Huang, Shikun Feng, Hui Zhong, Wenjin Wang, and Yu~Sun.
\newblock Masked label prediction: Unified message passing model for
  semi-supervised classification.
\newblock {\em arXiv preprint arXiv:2009.03509}, 2020.

\bibitem{spinoza2019reconstruction}
Hannah Spinoza and Douglas~B West.
\newblock Reconstruction from the deck of-vertex induced subgraphs.
\newblock {\em Journal of Graph Theory}, 2019.

\bibitem{stokes2020deep}
Jonathan~M Stokes, Kevin Yang, Kyle Swanson, Wengong Jin, Andres Cubillos-Ruiz,
  Nina~M Donghia, Craig~R MacNair, Shawn French, Lindsey~A Carfrae, Zohar
  Bloom-Ackermann, et~al.
\newblock A deep learning approach to antibiotic discovery.
\newblock {\em Cell}, 180(4):688--702, 2020.

\bibitem{tsitsulin2018netlsd}
Anton Tsitsulin, Davide Mottin, Panagiotis Karras, Alexander Bronstein, and
  Emmanuel M{\"u}ller.
\newblock Netlsd: hearing the shape of a graph.
\newblock In {\em Proceedings of the 24th ACM SIGKDD International Conference
  on Knowledge Discovery \& Data Mining}, pages 2347--2356, 2018.

\bibitem{ulam1960collection}
Stanislaw~M Ulam.
\newblock A collection of mathematical problems.
\newblock {\em New York}, 29, 1960.

\bibitem{velivckovic2017graph}
Petar Veli{\v{c}}kovi{\'c}, Guillem Cucurull, Arantxa Casanova, Adriana Romero,
  Pietro Lio, and Yoshua Bengio.
\newblock Graph attention networks.
\newblock {\em arXiv preprint arXiv:1710.10903}, 2017.

\bibitem{verma2017hunt}
Saurabh Verma and Zhi-Li Zhang.
\newblock Hunt for the unique, stable, sparse and fast feature learning on
  graphs.
\newblock {\em Advances in Neural Information Processing Systems}, 30, 2017.

\bibitem{wu2020graph}
Shiwen Wu, Fei Sun, Wentao Zhang, Xu~Xie, and Bin Cui.
\newblock Graph neural networks in recommender systems: a survey.
\newblock {\em ACM Computing Surveys (CSUR)}, 2020.

\bibitem{wu2020comprehensive}
Zonghan Wu, Shirui Pan, Fengwen Chen, Guodong Long, Chengqi Zhang, and S~Yu
  Philip.
\newblock A comprehensive survey on graph neural networks.
\newblock {\em IEEE transactions on neural networks and learning systems},
  2020.

\bibitem{xu2018powerful}
Keyulu Xu, Weihua Hu, Jure Leskovec, and Stefanie Jegelka.
\newblock How powerful are graph neural networks?
\newblock {\em arXiv arXiv:1810.00826}, 2018.

\bibitem{ying2021transformers}
Chengxuan Ying, Tianle Cai, Shengjie Luo, Shuxin Zheng, Guolin Ke, Di~He,
  Yanming Shen, and Tie-Yan Liu.
\newblock Do transformers really perform badly for graph representation?
\newblock {\em Advances in Neural Information Processing Systems},
  34:28877--28888, 2021.

\bibitem{zhang2018end}
Muhan Zhang, Zhicheng Cui, Marion Neumann, and Yixin Chen.
\newblock An end-to-end deep learning architecture for graph classification.
\newblock In {\em Proceedings of the AAAI conference on artificial
  intelligence}, volume~32, 2018.

\bibitem{zhu2019robust}
Dingyuan Zhu, Ziwei Zhang, Peng Cui, and Wenwu Zhu.
\newblock Robust graph convolutional networks against adversarial attacks.
\newblock In {\em Proceedings of the 25th ACM SIGKDD international conference
  on knowledge discovery \& data mining}, pages 1399--1407, 2019.

\bibitem{zugner2018adversarial}
Daniel Z{\"u}gner, Amir Akbarnejad, and Stephan G{\"u}nnemann.
\newblock Adversarial attacks on neural networks for graph data.
\newblock In {\em ACM SIGKDD}, 2018.

\end{thebibliography}

\end{document}